\documentclass[fleqn,a4paper]{article}

\usepackage{graphicx}  % standard LaTeX graphics tool
\usepackage{multicol}  % used for the two-column index
\usepackage{rotating}

\begin{document}
\title{Partial-wave decomposition 
of the Faddeev amplitude
of a Poincar\'e-invariant
three-fermion bound state\footnote{Talk given by R.A.\ at the
 43rd International Winter School on Theoretical Physics,
Schladming, Styria, Austria, February 26 - March 4, 2005}}

\author{Reinhard Alkofer\footnote{
Present address: Institute of Physics, Graz University, 
Universit\"atsplatz 5,
A-8010 Graz, Austria}\\
{\footnotesize Institute of Theoretical Physics, U.\ T\"ubingen, 
%Auf der Morgenstelle 14,
D-72070 T\"ubingen, Germany}\\[2mm]
and\\[2mm]
Martin Oettel\\
{\footnotesize 
Max-Planck Institute for Metal Research, %PO Box 80 06 65,
 D-70506 Stuttgart,
Germany}\\}

\maketitle              % typesets the title of the contribution

\section{Motivation}

The composite nature of the nucleon can be considered as the main source
of motivation to study of the relativistic 
three-fermion bound state problem. On the other hand, this  compositeness is 
also the source of our still 
incomplete understanding of the nucleon structure.
Although protons and neutrons (together with electrons) form the
building blocks of matter most of their properties are poorly understood.  {\it
E.g.\/} the experimental results on the spin structure of the proton have been
so surprising to model builders that they named the problem of explaining them
the {\em proton spin crisis}, see \cite{EL,SF} and references therein.

The aim of this lecture is much more modest than trying to explain the proton's
substructure in terms of quarks and gluons. Taking the simplest relativistic
and also realistic approach to model nucleons, namely, a Poincar{\'e} covariant
Faddeev approach to describe the binding of three valence quarks,  
we will demonstrate that even with relativistic valence quarks only 
the nucleon has quite a rich structure embodied in its
wave function. This will be exemplified in the nucleon's rest frame
by a decomposition into partial waves w.r.t.\  the motion of one of the
valence quark relative to the complementary  pair of quarks. 
As will be seen  this analysis (without referring to a specific dynamical model) 
also answers the question whether the nucleon is 
spherically symmetric: It is
not -- due to the highly relativistic motion of quarks within the nucleon.

\section{Spin of elementary particles}

To the best of our knowledge quarks are pointlike Dirac fermions with spin 
$\frac{1}{2}$.
It will prove helpful for the following to recall a few facts about their
relativistic description based on the solution of Dirac's equation.  First, Dirac
wavefunctions are four-component spinors. The physical reason for this is the
simultaneous description of particles and antiparticles. 
These four-component spinors are reduced to two components by a
projection onto positive energy states yielding a formalism akin to 
Pauli's equation.
Second, in the rest frame (in standard
representation) the lower two components vanish. Third, the upper
(``nonrelativistic") and the lower (``relativistic")
components carry different angular momentum, {\it e.g.\/} the lower components
represent a $p$-wave if the upper component corresponds to an $s$-wave. 
Fourth, interactions between the fermions can be incorporated unambiguously, 
the prime example is the causal coupling to the electromagnetic field.

Coupling the three spin--$\frac{1}{2}$ quarks to a composite spin--$\frac{1}{2}$ 
nucleon such that Poincar{\'e} covariance is
maintained we will see that \cite{Oettel:1998bk,Oettel:1999gc,Oettel:2000ig}:
\begin{itemize}
\item due to the compositeness, we need more components than four in total 
or two for the positive energy states.
\item the lower components will not vanish in the rest frame 
thus giving rise to the unavoidable presence of  at least a (relativistic)
$p$-wave contribution.
\item the difference of one angular momentum unit between upper and lower
components remains, however, there will be also a $d$-wave contribution.
\item the coupling to the electromagnetic field can be chosen such to maintain 
causality, however, at the expense of a fairly complicated structure
of the nucleon--photon vertex containing one-- and  two-loop contributions.
\end{itemize} 

As a remark we want to mention that three quarks may couple to form a total 
spin $\frac 3
2$, and, of course, they should do so to form the $\Delta$ baryon.  However,
already for elementary spin--$\frac 3 2$ objects, the Rarita-Schwinger fields, there
are a  number of problems: First, the corresponding spinors have sixteen components
but only eight of them are physical. 
Second, a ``non-relativistic'' limit does not
exist. Third, (if the field is not part of a supergravity multiplet)
interactions are not well-defined, {\it e.g.\/} the coupling to the
electromagnetic field is not causal. From this point of view it may appear 
surprising that within the Poincar{\'e} covariant Faddeev approach 
$\Delta$ baryons can be described \cite{Oettel:1998bk,Oettel:2000ig} -- 
which can be understood, however, from the finite extension of these composite
objects.

\section{Relativistic angular momentum:\\ Pauli-Lubanski vector}

In a nonrelativistic setting angular momentum is defined w.r.t.\ a fixed origin.
This makes plain why the concept of angular momentum has to be generalized 
properly in a
relativistic setting. Formally one sees the effect of relativity 
from the fact that the
the Casimir operator of the (non-relativistic) rotation group, $\vec J^2$, 
does not commute with boosts. 

Describing the angular momentum with the help of a vector operator
orthogonal to the  particle momentum will cure the underlying problem. Thus we 
will start our considerations from the Pauli-Lubanski (axial-) vector:
\begin{equation}
W_\mu = - \frac 1 2 \epsilon _{\mu\nu\rho\sigma } J^{\mu\nu} P^\sigma
\end{equation}
where  $J^{\mu\nu}$ is the Noether charge of rotations and boosts.
We note that
\begin{equation}
C_2=W_\mu W^\mu = m^2 j(j+1)
\end{equation}
is a (second) Casimir invariant of the Poincar\'e group, and that in the rest
frame it reduces to a quantity proportional to the usual spin:
\begin{equation}
W_\mu = (0,\vec W), \qquad W_i=- \frac 1 2 \epsilon _{ijk0} J^{jk} P^0 
= - m \Sigma _i \; .
\end{equation}

\section{Three-fermion states:\\ Partial wave decomposition in rest frame}

For the problem at hand we can reduce the complexity by noting that in a baryon
every quark pair is in a colour antitriplet state, and that the corresponding
interaction between the quarks is attractive.
(This can be surmised by calculating the group--theoretical factors for
a one--gluon exchange diagram; additionally, the attraction has been 
corroborated by lattice calculations.)
 Considering only states with vanishing orbital
angular momentum these quark pairs form scalar (spin 0) and axialvector (spin 1)
``diquarks''. Furthermore, the Pauli principle  requires flavour antisymmetry
for scalar and flavour symmetry for axialvector ``diquarks''.

\begin{figure}
\begin{center}
\includegraphics[width=.8\textwidth]{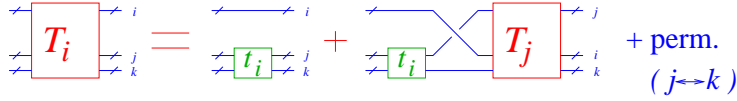}
\end{center}
\caption[]{Pictorial representation of the Poincar\'e covariant Faddeev
equation.}
\label{Faddeev}
\end{figure}

To obtain a Poincar\'e covariant Faddeev equation we consider Dyson's equation
for the 6-point function and neglect genuinely irreducible three-quark
interactions in its kernel. This leads to the equation depicted diagrammatically
in Fig.\  \ref{Faddeev}. 

\begin{figure}
\begin{center}
\includegraphics[width=.6\textwidth]{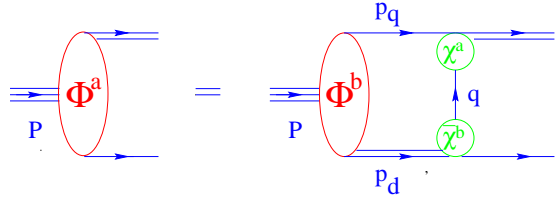}
\end{center}
\caption[]{Pictorial representation of the coupled quark-diquark
Bethe--Salpeter equations.}
\label{bse}
\end{figure}

By assuming that the two-body $t$-matrix can be approximated well by a finite
number of  separable quark--quark correlations\footnote{A rigorous separability 
expansion generates actually an infinite number of terms out of which we 
consider the (presumably) dominant ones.} into  the Poincar\'e covariant
Faddeev equation is mapped to a set of coupled Bethe--Salpeter equations. Its
structure is diagrammatically represented in Fig.\ \ref{bse}.  The
corresponding interaction is quark exchange. This reinstates the Pauli
principle at the level of all three valence quarks. As the colour quantum
number is antisymmetric, and thus all other quantum numbers are symmetric, 
the correlations induced by the Pauli principle amount to an attractive
interaction (as confirmed by the group--theoretical factor in  Fig.\ \ref{bse}). 

At this point we note that through our quite drastically simplifying assumptions
we have reduced the number of wave function components for the composite
nucleon from $4^3=64$ to still $4\cdot(1_{\rm sc.dq.}+3_{\rm ax.dq.})=16$ 
components.
A projection onto
positive energy reduces the  number of components to eight \cite{Oettel:1998bk}
which can be grouped into a set of  four bispinors consisting of an   upper and a 
lower component similar to Dirac spinors. 

\begin{figure}
\begin{center}
\includegraphics[width=.5\textwidth]{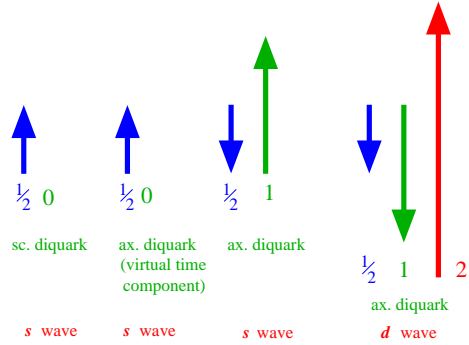}
\end{center}
\caption[]{A pictorial representation of the 
decomposition of the upper components of the nucleon's bispinors.}
\label{partial}
\end{figure}

To assess their physical properties we will first present a partial wave
decomposition in the rest frame. Note that due to the fact that neither orbital
angular momentum nor spin are good quantum numbers independent of the frame
the  following discussion is specific to this frame. First, we
decompose the Pauli-Lubanski vector into orbital and spin part
\begin{eqnarray}
W_i&=& {L_i + S_i}\,
  =\,\frac{1}{2} \epsilon_{ijk}\,{(\, {\cal L}^{jk}\,
    +\, {\cal S}^{jk}\,)} \\
{ {\cal L}^{jk}} &=& {\left( p^j
\frac{\displaystyle \partial }
    {\displaystyle  \partial p^k} - p^k \frac{\displaystyle \partial }
    {\displaystyle \partial p^j} \right)} \; , \\
{{\cal S}^{jk}} &=& { \frac{ 1}{ 2} \sigma^{jk}
   \otimes {\bf 1} \otimes {\bf 1} }\,\, + \,\, {\rm permutations.}
\end{eqnarray}
Here, $p$ is the relative momentim between the quark and the ``diquark" 
composite.
Second, we apply these operators onto the Bethe--Salpeter wave functions
\begin{eqnarray}
{ L^i L^i} \Phi_{\alpha\beta\gamma} &=&
   {l(l+1)} \Phi_{\alpha\beta\gamma} \; ,\\
 { S^i S^i} \Phi_{\alpha\beta\gamma} &=&
   {s(s+1)} \Phi_{\alpha\beta\gamma} \; ,
\end{eqnarray}
to obtain for the upper components three $s$-waves and one $d$-wave,
the lower components being four $p$-waves. A pictorial representation of the 
decomposition of the upper components is given in Fig.\ \ref{partial}.

\begin{table}
\caption{The eight components of the nucleon's Faddeev amplitude derived
with the simplifications described in the text, given as trispinors
$\Phi_{\alpha\beta\gamma}=U^i_\alpha(\gamma^iC)_{\beta\gamma}$. Scalar
diquark correlations correspond to $i=5$, axialvector ones to $i\equiv
\mu=1\dots 4$.}
\label{t}
\begin{center}
\begin{minipage}{1cm}
  \begin{sideways} scalar \end{sideways} \\
  \begin{sideways} \phantom{aaaaaaaaaaaa} \end{sideways} \\
   \begin{sideways} axialvector \end{sideways}
\end{minipage}
\hspace{-0.5cm}
\begin{tabular}{lcc} 
\hline \\
 nucleon wave function components  & eigenvalue & eigenvalue \\
 in the rest frame   & $l(l+1)$  & $s(s+1)$ \\
  &  of ${\bf L}^2$ & of ${\bf S}^2$ \\ \\  \hline \hline
  & & \\
{${\cal S}_1 u (\gamma_5 C)= {\chi \choose 0 }(\gamma_5 C)$}
 & 0 {s} & $\frac{3}{4}$ \\
\\
% ~~~~~~~~~~~~~~~~~~~~~~~~~~~~~~{scalar}&& \\
{${\cal S}_2 u (\gamma_5 C)={0\choose
\frac{1}{p}(\vec{\sigma}\vec{p})\chi }(\gamma_5 C)$}
 & 2 {p}& $\frac{3}{4}$ \\
 ~&&\\ ~&&\\
{${\cal A}^\mu_{1} u (\gamma^\mu C)=\hat
P^0{\frac{1}{p}(\vec{\sigma}\vec{p})\chi\choose 0}(\gamma^4 C)$}
 & 2 {p} &$\frac{3}{4}$\\
 &&\\
{${\cal A}^\mu_{2} u (\gamma^\mu C)=\hat P^0{0\choose
\chi}(\gamma^4 C)$}
 & 0 {s} & $\frac{3}{4}$ \\
 & & \\
{${\cal B}^\mu_{1} u (\gamma^\mu C)={i\sigma^i\chi \choose 0}(\gamma^i C)$}
 & 0 {s} & $\frac{3}{4}$ \\
\\
% ~~~~~~~~~~~~~~~~~~~~~~~~~{axialvector}&&\\
{${\cal B}^\mu_{2} u (\gamma^\mu
C)={0\choose\frac{i}{p}\sigma^i(\vec{\sigma}\vec{p})\chi}(\gamma^i C)$}
 & 2 {p} & $\frac{3}{4}$ \\
 & & \\
{${\cal C}^\mu_{1} u (\gamma^\mu C)={i\left(\hat
p^i(\vec{\sigma}\hat{\vec{p}})-\frac{1}{3}\sigma^i\right)\chi\choose
         0}(\gamma^i C)$}
 & 6 {d}& $\frac{15}{4}$ \\
 &&\\
{${\cal C}^\mu_{2} u (\gamma^\mu C)=
{0\choose \frac{i}{p}\left(p^i-\frac{1}{3}\sigma^i(\vec{\sigma}\vec{p})\right)
\chi}
               (\gamma^i C)$}
          & 2 {p} & $\frac{15}{4}$ \\
  & & \\
\hline
\end{tabular}
\end{center}
\end{table}

\section{Nucleons: Poincar\'e-covariant amplitudes}

In the Bethe--Salpeter
wave functions there appear two independent momenta, the total (nucleon) momentum
$P$
and the quark-diquark relative momentum $p$. Even in the rest frame the lower
components are non-vanishing. Similar as in Dirac's case the lower components
carry Pauli spin matrices but now these are contracted with the relative
momentum, see Table \ref{t}.
Thus we may  summarize shortly: A minimum of eight components is needed to describe 
the nucleon as  Poincar\'e covariant bound state of three quarks. The lower
components of this bispinorial quantities do NOT vanish, even not in the rest
frame. From this we conclude that if no mysterious cancelations occur the
nucleon is a non-spherical, deformed object.\footnote{Rotational symmetry is of
course guaranteed by the fact that all possible orientations are energetically 
degenerate, {\it cf.\/} the physics of deformed nuclei.}

The construction of the electromagnetic coupling of the nucleon within this 
approach can be found in the literature, see {\it e.g.\/} Refs.\ 
\cite{Oettel:1999gc,Kvinikhidze:1999xp}. Hereby the photon does not only couple
to the nucleons' constituents but also to the exchange quark and the
quark-diquark vertex functions. The corresponding terms can be either derived 
from the electromagnetic Ward identity \cite{Oettel:1999gc} or one employs the 
gauging-of-equations method \cite{Kvinikhidze:1999xp} which guarantees the
validity of the Ward identity from the beginning.

Calculations of the nucleons'
electromagnetic form factors employing these amplitudes and the consistent
coupling of the photon are in agreement with 
experimental data for larger momentum transfers
\cite{Oettel:2002wf,Alkofer:2004yf,Holl:2005zi}.  A failure at lower $Q^2$ 
had to be expected as
mesonic contributions are not present in this description. 
However, as mesons are composite objects their contribution dies out fast
for $Q^2$ above one to two GeV$^2$. 
A prediction of Ref.\ \cite{Holl:2005zi} is a zero of the
proton's electric form factor at $Q^2 \approx 8$\,GeV$^2$. In this  calculation
this zero can be traced back to the above described spinorial structure. 
Thus its
experimental verification or falsification may constitute a test of the 
considerations given above.

\section*{Acknowledgements}
Support by a
grant from the Ministry of Science, Research and the Arts of
Baden-W\"urttemberg (Az: 24-7532.23-19-18/1 and 24-7532.23-19-18/2)
is acknowledged.

\end{document}